\documentclass[a4paper]{article}
\usepackage{cite}
\usepackage{xcolor}
\usepackage{float}
\usepackage{INTERSPEECH2020}
\usepackage{makecell}
\usepackage{multirow,balance}

\usepackage{array}
\newcommand{\PreserveBackslash}[1]{\let\temp=\\#1\let\\=\temp}
\newcolumntype{C}[1]{>{\PreserveBackslash\centering}p{#1}}

\title{Audio-visual Speaker Recognition with a Cross-modal Discriminative Network}
\name{Ruijie Tao$^1$, Rohan Kumar Das$^1$ and Haizhou Li$^{1,2}$}
\address{
  $^1$Department of Electrical and Computer Engineering, National University of Singapore, Singapore
  $^2$Machine Listening Lab, University of Bremen, Germany}
\email{ruijie.tao@u.nus.edu.sg, \{rohankd, haizhou.li\}@nus.edu.sg}

\begin{document}
\maketitle
\begin{abstract}
%The 2019 speaker recognition evaluation (SRE19) released Audio-visual speaker recognition (audio-visual SRE) task, which use both the audio and visual information to do speaker recognition.
Audio-visual speaker recognition is one of the tasks in the recent 2019 NIST speaker recognition evaluation (SRE). Studies in neuroscience and computer science all point to the fact that vision and auditory neural signals interact in the cognitive process. This motivated us to study a cross-modal network, namely voice-face discriminative network (VFNet) that establishes the general relation between human voice and face. Experiments show that VFNet provides additional speaker discriminative information. With VFNet, we achieve 16.54\% equal error rate relative reduction over the score level fusion audio-visual baseline on evaluation set of 2019 NIST SRE.

%In this paper, inspired by the experiments in neuroscience and computer science, which has demonstrated there are some overlapped speaker verification related information between human voices and faces. We propose a novel cross-modal verification network: Voice-face network (VFNet), which can achieve state-of-the-art cross-modal verification/matching results. Secondly, we focus on the application value of introducing the cross-modal information in practical speaker recognition scenario. We apply VFNet in audio-visual SRE, use cross-modal recognition system to assist the original audio-visual SRE framework. Compared with the baseline audio-visual SRE framework, the proposed method's EER, minDCF and actDCF can improve by 15.48\%, 12.32\% and 7.94\% in the evaluation set.

\end{abstract}

\vspace{2mm}


\section{Introduction}
% SV, SRE, SRE19
Speaker recognition has enabled many real-world applications~\cite{lee2011joint,sv_debut,das2017development,SpeechMarker}. These systems are expected to perform effectively under adverse conditions. The NIST speaker recognition evaluation (SRE)s are organized to benchmark systems in different such scenarios~\cite{greenberg2020two}.
%to spearhead the community~\cite{greenberg2020two}. 
Various robust systems are developed in the past that perform effectively and provides state-of-the-art~\cite{Villalba2019,I4U2019}. Unlike previous SREs, the 2019 NIST SRE investigated a new direction on audio-visual (AV) SRE~\cite{sadjadi20202019}. The evaluation task deals with verifying the claimed identity of a person for a given pair of enrollment and test videos. In other words, it advocates the use of audio-visual cues for improved speaker recognition in real-world scenarios.
%research along such direction can lead towards robust systems for practical scenarios.

% Multimedia audio-visual application in speech area (except SV)
The significance of processing multimedia in other fields has increased in the recent years~\cite{baltruvsaitis2018multimodal}. The latest audio-visual SRE can be viewed as one such outcome following this trend. Some of the other tasks considering multimedia instead of single modality using speech are automatic speech recognition (ASR)~\cite{noda2015audio}, speech separation~\cite{ephrat2018looking} and speech diarization~\cite{hoover2017putting}. The studies in these works exploited the association of audio and visual cues adequately. For instance, in audio-visual ASR, lip language recognition is used to support ASR systems; in audio-visual speech separation, the movement of mouth can assist detecting who is speaking when.

% Challenge and motivation
While coming to audio-visual SRE, the simplest way to perform multimedia based speaker recognition is to have separate systems for audio and visual inputs, then combine the results of speaker and face recognition systems~\cite{sadjadi20202019,HLT_SRE2019}. We note separating audio-visual SRE into two sub-tasks is a straight forward approach to simplify the problem. However, the two subsystems are disjoint and one does not consider the knowledge from other. It is further worth emphasizing that the motivation to process multimedia for SRE is not only to add another visual system but also to explore the relationship between the audio and video. Therefore, disregarding the association between different modalities may result in the loss of some information.

% Face-voice relationship in SV
%Is there any relationship between voice and face can explore for verification? 

The studies in neuroscience show that humans associate the voice and the face of a person in the memory~\cite{kriegstein2005interaction}. While listening to a voice of an individual, one can select the right static face corresponding to the same person between two static faces at a higher than chance level and vice versa~\cite{kamachi2003putting, mavica2013matching, smith2016matching, smith2016concordant}. In computer science, there has been study on cross-modal biometric matching~\cite{nagrani2018seeing}. Further, various works use pair-wise loss-based methods to improve the cross-modal system performance~\cite{horiguchi2018face, xiong2019voice}. The learned associations between audio and visual cues are general identity features (such as gender, age and ethnicity) and appearance features (such as big nose, chubby and double chin)~\cite{kim2018learning, oh2019speech2face}. The existing works utilize both joint and disjoint general information between two modalities to train the cross-modal verification network to determine if the given face and speech segment belongs to the same identity for verification tasks~\cite{nagrani2018learnable, wen2018disjoint, nawaz2019deep}. We believe that the general cross-modal discriminative features provide additional information in audio-visual speaker recognition.

%the cross-modal verification systems have potential scope for assisting audio-visual SRE by providing associative audio-visual information.

%there is still a large room to optimize the cross-modal verification system and reveal its practical value.

In this work, we propose a cross-modal discriminative network, that is called voice-face network (VFNet), to learn the association between voice and face. 
%a voice-face network (VFNet) based cross-modal discrimination network for establishing the relation between audio and video. The 
VFNet is trained using speaker and face embeddings collectively that are extracted from separate systems. We consider x-vector and InsightFace based systems for extracting the speaker and face embeddings, respectively~\cite{snyder2018x,deng2019arcface}. Further, these two systems are used for SRE using audio as well as visual input based single systems, followed by their fusion for a baseline audio-visual system. The output of VFNet is used to represent the general association between voice and face. The speaker recognition studies are conducted on 2019 NIST SRE corpus. The contributions of this paper include the novel idea of cross-modal discriminative network, and its use in audio-visual speaker recognition study.  

%The contribution of this work lies in proposal of a novel VFNet based cross-modal discrimination network, followed using it for audio-visual SRE studies. To the best of our knowledge, no previous work used cross-modal discrimination network in the context of multimedia speaker recognition.   

%According to that, our goal is to build a robust, reasonable and reliable cross-modal verification system to mine more voice-face information. Import this system to do cross-modal recognition in audio-visual SRE, to show the application value of cross-modal model in speaker recognition task. In this paper, we make the following contributions:
% What we did 
%\begin{itemize}
%\item We propose Voice-face network (VFNet) for cross-modal verification. Different from existed works, VFNet applied the verification related embeddings and trained on novel VoxCeleb2 database, state-of-the-art results can be achieved.
%\item We introduce VFNet into audio-visual SRE. Utilize the cross-modal recognition to assist the face recognition and speaker recognition. To the best of our knowledge, this work is the first one to use the cross-modal (between face and voice) system in speaker recognition and face recognition. We experimentally verify that cross-modal recognition can improve the audio-visual SRE results.
%\end{itemize}

% Structure of the rest paper
The rest of the paper is organized as follows. Section~\ref{secii} describes the proposed VFNet based cross-modal verification system. In Section~\ref{seciii}, we present the audio-visual speaker recognition with  cross-modal verification. Section~\ref{seciv} and Section~\ref{secv} reports the experiments and their results, respectively. %Section~\ref{secv} reports the results and analysis for VFNet and VFNet aided audio-visual SRE.
Finally, Section~\ref{conc} concludes the work.

\section{Cross-modal Verification}
\label{secii}

\begin{figure}[t]
  \centering
  \includegraphics[width=\linewidth]{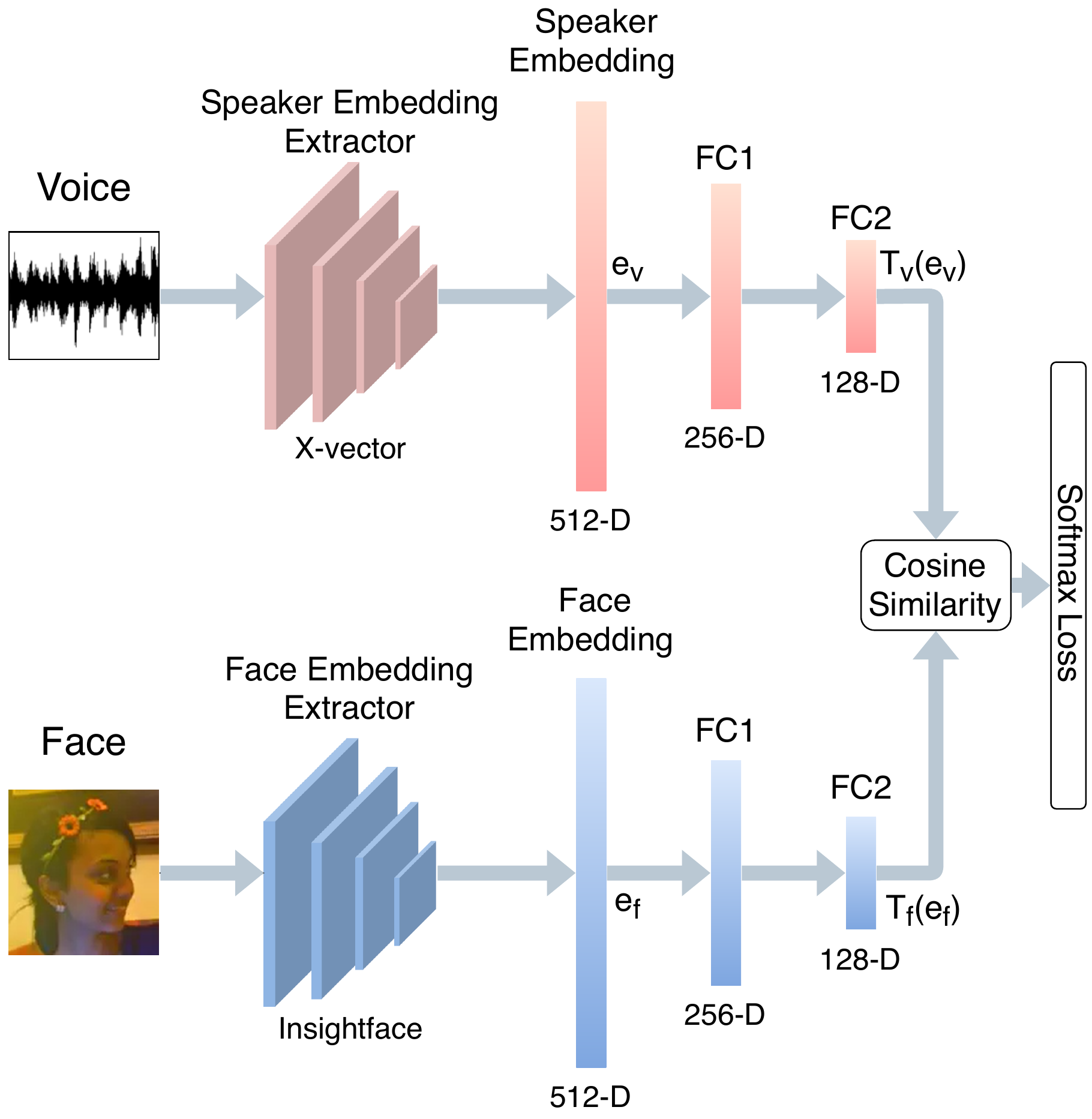}
  \caption{Architecture of the proposed cross-modal discrimination network, VFNet, that relates the voice and face of a person.}
  %\caption{The overall of the proposed cross-modal verification system: Voice-face Network (VFNet). The figure describes the training process for each trial.}
  \label{fig:VFNet}
\end{figure}

This section describes the proposed VFNet for cross-modal verification. Figure~\ref{fig:VFNet} shows the architecture of VFNet that considers two inputs: a voice waveform and a human face. The output of the network is a confidence score to describe whether the voice and the face come from the same person. We now discuss the detailed pipeline of VFNet.

First, the speaker and the face embeddings are extracted by x-vector and InsightFace models, respectively~\cite{snyder2018x,deng2019arcface}. As both these embeddings represent information from different modality, they are fed to a 256-D fully connected layer (FC1) with the rectified linear unit (ReLU) activation, then followed by another 128-D fully connected layer (FC2) without the ReLU. These layers are introduced to lead the speaker and face embeddings for learning the cross-modal identity information from each other. Further, they help to project the embeddings from both modalities into a new domain, where their relation can be established.
%The network can induce them to learn the overlapped information from each other in the training process. Accordingly, they can be transferred into the similar information format embeddings automatically. The role of these layers is to lead the speaker and face embeddings learning the overlapped identity related information from each other. Accordingly, they can be transferred into the similar new kinds of embeddings, which can be used to calculate the cross-similarity.

%Firstly, we propose our cross-modal verification model: Voice-face Network (VFNet), as shown in Figure~\ref{fig:VFNet}. The input is a voice waveform and a human face. The output is the metric to describe whether the voice and the face come from the same person. Obtain speaker and face embeddings by x-vector extractor and Insightface model respectively, followed by cross-modal system to calculate the verification metric based on these embeddings~\cite{snyder2018x,deng2019arcface}.

%The extracted speaker and face embeddings are different kinds of embeddings, feed each of them to a 256-D fully connected layer with the ReLU activation, then followed by another 128-D fully connected layer without the ReLU. The network can induce them to learn the overlapped information from each other in the training process. Accordingly, they can be transferred into the similar information format embeddings automatically.

For a given pair of speaker embedding $e_v$ and a face embedding $e_f$, their transformed embeddings $T_v(e_v)$ and $T_f(e_f)$ are derived from VFNet,  followed by the cosine similarity scoring $S(T_v(e_v),T_f(e_f))$ between them. We also have $1 - S(T_v(e_v),T_f(e_f))$ to represent the negative voice-face pair. By using softmax function based on these two scores, the output of VFNet is obtained as
%$S(T_v(e_v),T_f(e_f))$ calculates the cosine similarity between the transferred speaker embedding $T_v(e_v)$ and face embedding $T_f(e_f)$ . Define the VFNet's output in Equation~(\ref{equ:output_V&F_belong_to_same_person, softmax}) and ~(\ref{equ:output_VF_belong_to_different_person, softmax}). The final output $p_1$ is the score to describe the probability that the voice and the face belong to the same person.
\begin{equation}
  p_1 = \frac{e^{S(T_v(e_v),T_f(e_f))}}{e^{S(T_v(e_v),T_f(e_f))} + e^{1-S(T_v(e_v),T_f(e_f))}}
  \label{equ:output_V&F_belong_to_same_person, softmax}
\end{equation}
\begin{equation}
  p_2 = \frac{ e^{1-S(T_v(e_v),T_f(e_f))}}{e^{S(T_v(e_v),T_f(e_f))} + e^{1-S(T_v(e_v),T_f(e_f))}}
  \label{equ:output_VF_belong_to_different_person, softmax}
\end{equation}
  where final output $p_1$ is the score to describe the probability that the voice and the face belong to the same person, $p_2$ being the score depicting the probability that the voice and the face do not belong to the same person. Finally, we feed our predictions $p$ and the ground truth verification labels $\hat{p}$ to optimize the cross-entropy loss $\mathcal{L}_{\hat{p}}(p)$ as follows
\begin{equation}
  \mathcal{L}_{\hat{p}}(p) = -\sum_i{\hat{p_i}\log(p_i)}
  \label{equ:loss function, cross-entropy}
\end{equation}

%%%%%%%%%%%%%%%%%%%%%%%%%%%%%%%%%%%%%%%%%%%%%%%%%%%%%%%%%%%%%%%%%

\section{Audio-visual Speaker Recognition with Cross-modal Verification}
\label{seciii}

\begin{figure*}[t]
  \centering
  \includegraphics[width=1\linewidth]{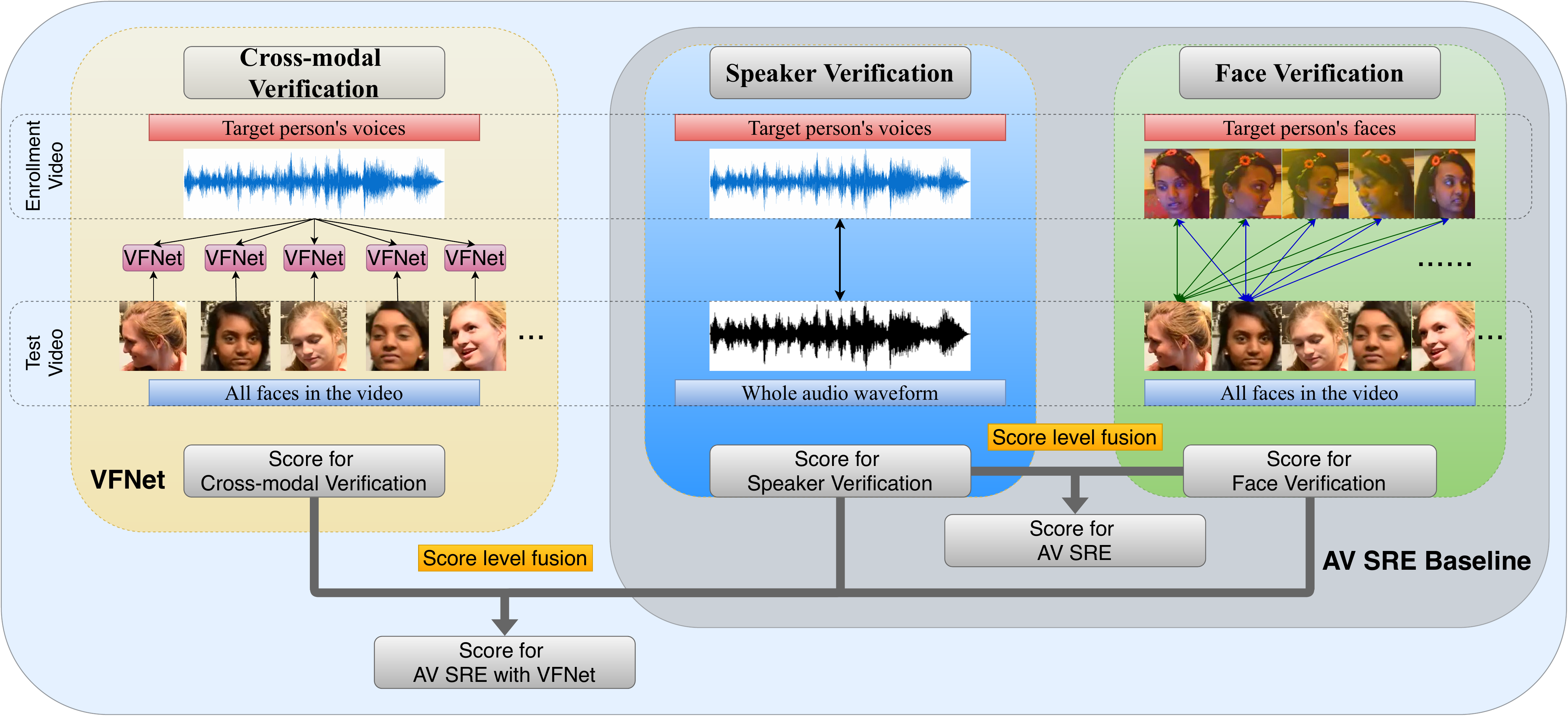}
  \vspace{2mm}
  \caption{Block diagram of proposed audio-visual (AV) speaker recognition framework with VFNet, where VFNet provides voice-face cross-modal verification information that strengthens the baseline audio-visual speaker recognition decision. }
  %\caption{The grey part: Audio-visual SRE baseline framework; the whole structure: proposed VFNet aided audio-visual SRE framework.}
  \label{fig:VFNet aided audio-visual SRE}
 % \vspace{-4mm}
\end{figure*}

In audio-visual SRE, an enrollment video provides the target individual's biometric information (voice and face) and the assignment asks the model to automatically determine whether the target person is present in a given test video~\cite{sadjadi20202019}.

Figure~\ref{fig:VFNet aided audio-visual SRE} shows the proposed  audio-visual speaker recognition framework with VFNet on the left panel, and the baseline, a voice-face score level fusion system, on the right panel. The given voice segments of the target speakers from the enrollment utterances and the entire test utterances are considered for extracting the speaker embeddings using x-vector system~\cite{snyder2018x}. Similarly, the InsightFace system extracts the face embeddings for given faces of the target speakers from the enrollment videos and all detected faces from the test videos~\cite{deng2019arcface}. 

On the left panel, the VFNet system provides an association score between the target speaker voice in the enrollment and the detected faces from the test video. Matching pairs between voice and face will give rise to high association, while mismatches, such as age, gender, and ethnicity discrepancy, will do otherwise.
  
On the right panel, the audio and visual systems run in parallel to verify the claimed identity by computing match between the enrollment and the test embeddings. We note that the speaker recognition system considers probabilistic linear discriminant (PLDA) based likelihood scores, whereas the face recognition system computes cosine similarity scores. We consider the baseline system as a score level fusion between the two parallel systems.

We propose to fuse the VFNet score and the baseline audio-visual system as shown in Figure~\ref{fig:VFNet aided audio-visual SRE} for a final decision. The score level fusion is performed using logistic regression for various systems discussed in this work. We report performance of the VFNet, baseline systems, and the overall system separately in the experiments.

%Learnt from the popular systems fusion method in speaker recognition, logistic regression can be applied to different systems in order to find the best detection cost function, which is the final recognition scores to evaluate the performance in SRE 2019 requirements. 

% The result from VFNet can also be used with the baseline audio-visual system as shown in Figure~\ref{fig:VFNet aided audio-visual SRE}. Considering speaker and face recognition systems are equally significant, the optimal weight for scores from cross-modal verification can be derived as given in Equation~(\ref{equ:fusion:baseline  score + VFNet score}) in the context of audio-visual SRE with VFNet. 
% \begin{equation}
%   \theta^{w} = \argmin_{\theta}E(s_s + s_f + \theta \times s_c) \\
%   \label{equ:fusion:baseline score + VFNet score}
% \end{equation}

\section{Experiments}
\label{seciv}
In this section, the details of the audio and visual systems developed in this work are mentioned. The database, embedding extraction and
audio-visual speaker recognition systems are described in the following subsections.

\subsection{Database}

We consider the original videos corresponding to VoxCeleb2 corpus to derive a set with voices and faces for cross-modal verification~\cite{chung2018voxceleb2}. For each video, the entire audio is extracted to represent the voice of the speaker. On the contrary, we perform a face detection on each video and then consider the most prominent faces representing an individual. For cross-modal discriminative training, the positive trials are faces and voices come from the same identity, whereas the negative trials are obtained by shuffling the faces and voices belonging to different persons. A summary of VoxCeleb2 corpus used for cross-modal verification is shown in Table~\ref{tab:VoxCeleb2}. VFNet learns the general association between voice and face from VoxCeleb2.

%Cross-modal related database needs to be built for VFNet training and evaluation. One method is to find the overlapped identities in one speaker recognition database and one face recognition database, combine their voices and faces to build the cross-modal database. But it might lead to the voice and face come from the same identity's different periods. To avoid that, we select voices and faces from the same segments in VoxCeleb2 database (video format)~\cite{chung2018voxceleb2}. In each video, extract the whole waveform to present the individual's voice information; obtain the first frame, do face detection and cut out the most prominent face to perform the individual's face information. Positive trials are faces and voices come from the same identity, randomly shuffle and recombine them to create the same number of negative trials. The statistics of the modified VoxCeleb2 database are tabulated in Table~\ref{tab:VoxCeleb2}.

\begin{table}[t]
  \caption{Summary of VoxCeleb2 and 2019 SRE audio-visual (AV) corpora.}
  %\vspace{-2mm}
  %\caption{Statistics of the modified VoxCeleb2 database used for cross-modal verification and the SRE19 (AV) database for audio-visual SRE.}
  \label{tab:VoxCeleb2}
  \centering
  \setlength{\tabcolsep}{2mm}{
  \begin{tabular}{ c@{}c c c}
    \toprule
    \multicolumn{2}{c}{\textbf{VoxCeleb2}} & \textbf{Train} & \textbf{Test}\\
    \midrule
    \multicolumn{2}{c}{\# identity} & $5,994$ & $108$~~~\\
    \multicolumn{2}{c}{\# faces} & $1,088,047$ & $36,166$~~~\\
    \multicolumn{2}{c}{\# voices}& $1,092,009$ & $36,237$~~~\\
    \multicolumn{2}{c}{\# cross-modal trials}& $2,176,094$ & $72,332$~~~\\
    \midrule
    %\hline
    \multicolumn{2}{c}{\textbf{2019 SRE (AV)}} & \textbf{Development} & \textbf{Evaluation} \\
    \midrule
    \multicolumn{2}{c}{\# enroll segments}  & $52$ & $149$~~~ \\
    \multicolumn{2}{c}{\# test segments } & $108$ & $452$~~~ \\
    \multicolumn{2}{c}{\# target trials} & $108$ & $452$~~~ \\
    \multicolumn{2}{c}{\# non-target trials} & $5,508$ & $66,896$~~~ \\ 
    \bottomrule
  \end{tabular}}
  %\vspace{-4mm}
\end{table}	

%The 2019 NIST SRE audio-visual corpus is used for audio-visual speaker recognition studies

We consider 2019 NIST SRE audio-visual corpus as the speaker recognition application~\cite{sadjadi20202019}. The evaluation set has provided manually marked diarization labels for voice and keyframe indices along with target speaker's face bounding boxes in the enrollment videos. However, no such information is provided for the test segments. A summary of this corpus is also shown in Table~\ref{tab:VoxCeleb2}. We note that to maximize the usage of cross-modal information in this corpus, we extract the speaker and face embeddings of the target person from the enrollment segments of development set, then combine them with that of VoxCeleb2 to retrain the VFNet.

%In audio-visual SRE, NIST provides the SRE19 (AV) database to test the system's performance. SRE19 (AV) includes the development set and the evaluation set, each of them has the enrollment set and test set. To maximize the usage of cross-modal information in this database, we extract the target individual's speaker and face embeddings from the enrollment set, combine them into trials to retrain our VFNet again. The statistics of the SRE19 (AV) database are also in Table~\ref{tab:VoxCeleb2}.

\subsection{Embedding Extraction}

We use an x-vector based system to extract the speaker embeddings~\cite{snyder2018x}. The speech utterances are processed with an energy based voice activity detection to remove the non-silence regions and 30-dimensional mel frequency cepstral coefficient (MFCC) features are extracted. In addition,  a short-time cepstral mean normalization is applied over a 3-second sliding window. The x-vector extractor is trained using VoxCeleb1-2 corpora and the detailed settings of x-vector network architecture can be found in~\cite{snyder2019speaker}.

%In VFNet for speaker embedding extraction, different from the existing works which select the general neural networks to do the extraction, we extract embeddings based on x-vector, which can provide robust DNN embeddings for speaker recognition. Firstly, 30-dimensional Mel frequency cepstral coefficient (MFCC) are extracted from each utterance, before removing the non-speech frames by the energy-based voice activity detection (VAD), a short-time cepstral mean normalization was applied over a 3-second sliding window. followed by the x-vector extractor, which is trained on VoxCeleb 1 and 2, to obtain the speaker embeddings in Kaldi~\cite{povey2011kaldi}, the setting of x-vector network architecture can be found in~\cite{snyder2018x}.

%The face recognition system consists of two stages that are face detection and then their recognition.

For extracting the face embeddings, we first use the ResNet-50 RetinaFace model trained using WIDER FACE database~\cite{yang2016wider} for detecting the faces~\cite{deng2019retinaface} followed by multi-task cascaded convolutional network (MTCNN)~\cite{zhang2016joint} to align them. We then use InsightFace to obtain highly discriminative features for face recognition by using the additive angular margin loss~\cite{deng2019arcface}. In addition, it consists of ResNet-100 extractor model trained on cleaned MS1MV2 database~\cite{guo2016ms} to extract the face embeddings. 

The dimension of both speaker and face embeddings are kept as 512 in our studies. The speaker and face embeddings for VFNet to perform cross-modal verification follow the same pipeline discussed above. 

%For face embedding, the ResNet-50 RetinaFace model is trained in the WIDER FACE database~\cite{yang2016wider} to do face detection~\cite{deng2019retinaface}, then aligned by Multi-task Cascaded Convolutional Network (MTCNN)~\cite{zhang2016joint}. InsightFace is then used to obtain highly discriminative features for face recognition by using the additive angular margin loss~\cite{deng2019arcface}. The ResNet-100 extractor model was trained on the VGGFace2~\cite{cao2018vggface2} and cleaned MS1MV2 database~\cite{guo2016ms}. The dimensions of the speaker and face embedding are both 512.

\subsection{ Audio-visual Speaker Recognition}

Although the dimensions of speaker and face embeddings are the same, the back-end scoring for respective individual system is different. We use linear discriminant analysis (LDA) on speaker embeddings for channel/session compensation and reduce the dimension of x-vectors to 150. Finally, PLDA is used as a classifier to get the final speaker recognition score.  On the other hand, cosine similarity between face embeddings from enrollment video and that from the detected faces in the test video are computed. Finally, the average of top 20\% scores of the number of face embeddings in the test video are taken to derive the final face recognition score.  

%For speaker recognition, x-vector is utilized to obtain the embeddings, followed by linear discriminant analysis (LDA) to minimize the channel and distance variation and reduce the dimension of x-vectors to 150. Out-of-domain PLDA is utilized as the classifier to get the final speaker recognition score.

%In the face recognition system, for each trial, extract $N_e$ and $N_t$ face embeddings from enrollment and test video by the same face framework in VFNet. Calculate the cosine similarity between them, apply the average of the highest $N$ similarity scores in the obtained score array (where $N = 20\% \times N_t$) to perform the face recognition result.

%For the VFNet based cross-modal recognition system in audio-visual SRE. The back-end is to calculate the average of the highest $20\%$ scores in all scores from VFNet. This result describes the probability that there is someone in the test video can make the target person's sound. 

We now focus on the back-end of audio-visual SRE with VFNet. The VFNet back-end computes the likelihood score between the speaker embedding of the target speaker and all the face embeddings of detected faces in the test video as given by Equation~(\ref{equ:output_V&F_belong_to_same_person, softmax}). Finally, the average of top 20\% scores is taken, which is then combined with the scores generated from audio and visual systems by logistic regression. It is to be noted that cross-modal verification can also be done by considering the all the given faces in the enrollment video and the detected multiple speaker voices in the test audio. However, it requires an additional speaker diarization module to detect the voice belonging to different speakers in the test audio. Therefore, we follow the former approach for audio-visual SRE with VFNet.    

We use Bosaris toolkit~\cite{bosaris} to calibrate and fuse the scores of different systems. The performance of systems are reported in terms of equal error rate (EER), minimum detection cost function (minDCF) and actual detection cost function (actDCF) following the protocol of 2019 NIST SRE~\cite{sadjadi20202019}. 

%Using logistic regression to calibrate the speaker and face recognition results~\cite{brummer2007fusion}. Due to the independence of these two systems, add the calibrated scores up for the audio-visual SRE baseline results. According to multimedia SRE19's requirements, we reports the results in terms of equal error-rate (EER), minimum detection cost function (minDCF) and actual detection cost function (actDCF). Here actDCF is the primary metric.

\section{Results and Analysis}
\label{secv}

\subsection{Cross-modal Verification Studies}

We evaluate the performance of proposed VFNet on VoxCeleb2 corpus for cross-modal verification studies and compare with some of the existing systems. The performance comparison is shown in Table~\ref{tab:cross-modal verification results}, where EER and area under the ROC curve (AUC) are considered as the performance metrics. 

We observe that VFNet performs effectively for cross-modal verification. Further, it is to be noted that we do not claim from this study that VFNet outperforms other systems as the results are evaluated on different corpora. We rather try to show that VFNet alone is comparable to the existing systems for cross-modal verification. For brevity, we do not go through the details of various systems~\cite{nagrani2018learnable,nawaz2019deep,wen2018disjoint} considered for cross-modal verification.

%Firstly, Table~\ref{tab:cross-modal verification results} reports the VFNet's performance in cross-modal verification task. Standard metrics are applied for verification: area under the ROC curve (AUC) and EER. Due to the reasonable structure and the loss function of VFNet for cross-modal verification problem. VFNet can perform the comparable existing state-of-the-art AUC and EER results. In addition, x-vector and Insightface are more suitable to describe embeddings than other neural networks. Moreover, We extract the face and voice from the same video segment in the sizeable VoxCeleb2 database instead of stitching databases. 

\begin{table}[t]
  \caption{Comparative study between the proposed VFNet and  other systems in cross-modal verification.} 
 % \vspace{-2mm}
  %For the model using two databases, it means the overlapped identities' voice and face from these the speaker recognition database and the face recognition database are selected.}
  \label{tab:cross-modal verification results}
  \centering
  \setlength{\tabcolsep}{0.8mm}{
  \resizebox{8cm}{!}{
  \begin{tabular}{ c@{}c c c c}
    \toprule
    \multicolumn{2}{c}{\textbf{Model}} &\textbf{EER (\%)} & \textbf{AUC (\%)} &\textbf{Database}\\
    \midrule
    \multicolumn{2}{c}{DIMNet~\cite{wen2018disjoint}} & $24.56$ & NA & VoxCeleb, VGGFace~~~             \\
    \multicolumn{2}{c}{SSNet~\cite{nawaz2019deep}} & $29.5$ & $78.8$ & VoxCeleb~~~             \\
    \multicolumn{2}{c}{Pins~\cite{nagrani2018learnable}} & $29.6$ & $78.5$ & VoxCeleb~~~\\
    \multicolumn{2}{c}{\textbf{VFNet}} & $\textbf{22.52}$ & $\textbf{85.4}$ & VoxCeleb2~~~\\
    \bottomrule
  \end{tabular}}
  }
 % \vspace{-2mm}
\end{table}

%To measure VFNet's performance more comprehensively, we also do experiments to achieve the cross-modal matching task, which the existed cross-modal works focus more. This task provides a voices and two candidate static human faces, ask the model to select which face is more inclined to make this voice. We add one more face sub-branch to the original VFNet model, let two face sub-branches share weights then do experiments in modified VoxCeleb2 to deal with this task, similar network can be built if we use the face as the reference modality. The metric used is the accuracy. The results in Table~\ref{tab:cross-modal matching results} show VFNet can also perform the outstanding ACC results in cross-modal matching task. 

\begin{table}[t]
  \caption{Comparative study between the proposed VFNet and other systems in cross-modal matching. V-F and F-V refer to cases that consider reference modality as voice and face, respectively.}
  %\vspace{-2mm}
  %V-F means using the voice as the reference modality for matching and F-V means using the face as the reference.}
  \label{tab:cross-modal matching results}
  \centering
  \setlength{\tabcolsep}{1.15mm}{
  \resizebox{8cm}{!}{
  \begin{tabular}{ c@{}c c c c}
    \toprule
    \multicolumn{2}{c}{\multirow{2}{*}{\textbf{Model}}} & \multicolumn{2}{c}{\textbf{Accuracy (\%)}} & \multicolumn{1}{c}{\multirow{2}{*}{\textbf{Database}}} \\
    \multicolumn{2}{c}{~} & \textbf{V-F} & \textbf{F-V} & \multicolumn{1}{c}{~}  \\
    \midrule
    \multicolumn{2}{c}{SSNet~\cite{nawaz2019deep}}& $78$ & NA & VoxCeleb~~~\\
    \multicolumn{2}{c}{Horiguchi's~\cite{horiguchi2018face}} & $78.1$ & $77.8$ & VoxCeleb, MS-Celeb-1M ~~~\\
    \multicolumn{2}{c}{Kim's~\cite{kim2018learning}}& $78.2$ & $78.6$ & VoxCeleb~~~             \\
    \multicolumn{2}{c}{SVHF~\cite{nagrani2018seeing}} & $81.0$ & $79.5$ & VoxCeleb, VGGFace~~~\\
    \multicolumn{2}{c}{Pins~\cite{nagrani2018learnable}} & $84$ & NA & VoxCeleb~~~\\
    \multicolumn{2}{c}{DIMNet~\cite{wen2018disjoint}} & $84.12$ & $84.03$& VoxCeleb, VGGFace~~~\\
    \multicolumn{2}{c}{VFMR~\cite{xiong2019voice}} & $84.48$ & NA & VoxCeleb2, VGGFace2~~~\\ 
	\multicolumn{2}{c}{\textbf{VFNet}}& $\textbf{85.39}$ & $\textbf{86.12}$ & VoxCeleb2~~~\\
    \bottomrule
  \end{tabular}}
  } 
  %\vspace{-4mm}
\end{table}

Further, to measure VFNet performance more comprehensively, we extend the studies for cross-modal matching task. For a given human voice and two static faces, this task aims to find the more inclined face to the voice, and vice versa. We note that this task also relates to 2019 NIST audio-visual SRE as there are multiple speakers present in the test videos that have to be matched with the target speaker in the enrollment video. For this cross-modal matching study, we add one more shared weights sub-branch to the original VFNet model for the selection requirements. The performance of VFNet thus obtained and its comparison to some of the other systems for cross-modal matching task in terms of accuracy is shown in Table~\ref{tab:cross-modal matching results}. We observe that the effectiveness of VFNet holds good for cross-modal matching task as well and the performance is comparable to other systems.

%\textcolor{red}{Haizhou: I totally don't understand this paragraph. I cannot revise. Further, we extend the studies for cross-modal matching task that deals with finding out the more inclined face to a voice from two given faces. For this study, we add one more face sub-branch to the original VFNet model, where the two face sub-branches share weights for the studies. The performance of VFNet thus obtained and its comparison to some of the other systems for cross-modal matching task in terms of accuracy is shown in Table~\ref{tab:cross-modal matching results}. We observe that the effectiveness of VFNet holds good for cross-modal matching task as well and the performance is comparable to other systems.}

\subsection{Audio-visual SRE with VFNet Studies}

We now study audio-visual speaker recognition with VFNet. To show the effect of VFNet, we first fuse the single modality speaker, face recognition systems with VFNet. Table~\ref{tab:VFNet aided audio-visual SRE results} reports the performance comparison of various systems and with and without VFNet.

Examining the effect on single modality systems, we find that the contribution of VFNet is more evident for speaker recognition system on the evaluation set. Further, the VFNet is also able to enhance the audio-visual baseline system performance that suggests usefulness of associating audio and visual cues by cross-modal verification for audio-visual SRE. We obtain relative improvements of 16.54\%, 2.00\% and 8.83\% in terms of EER, minDCF and actDCF, respectively.

%In order to compare the performance between the baseline and the proposed methods, we report the baseline results in audio-visual SRE in Table~\ref{tab:VFNet aided audio-visual SRE results}. Speaker recognition (SR), face recognition (FR) and audio-visual fused baseline results are shown in it. We find the fusion of the SR and FR systems enhances the performance by a larger margin due to the complementary information for capturing speaker identity. 

\begin{table}[t]
  \caption{Performance comparison of various systems on 2019 NIST SRE audio-visual corpus.}
 % \vspace{-2mm}
  \label{tab:VFNet aided audio-visual SRE results}
  \centering
  \resizebox{8cm}{!}{
  \setlength{\tabcolsep}{0.8mm}{
  \begin{tabular}{ c@{}c C{13mm} C{12mm} C{12mm}}
    \toprule
    \multicolumn{5}{c}{\textbf{Development Set}} \\
    \midrule
    \multicolumn{2}{c}{\textbf{System}} & \textbf{EER (\%)} & \textbf{minDCF} & \textbf{actDCF}  \\
    \midrule    
    \multicolumn{2}{c}{Speaker Recognition} & $08.62$ & $0.367$ & $0.399$ ~~~             \\
    \multicolumn{2}{c}{with VFNet} & $09.82$ & $0.365$ & $0.393$ ~~~             \\
    \midrule
    % \multicolumn{2}{c}{{VFNet}} & $0.173$ & $0.196$ & $03.98$~~~             \\
    \multicolumn{2}{c}{Face Recognition } & $04.52$ & $0.349$ & $0.371$ ~~~             \\
    \multicolumn{2}{c}{with VFNet} & $03.85$ & $0.324$ & $0.355$ ~~~             \\
    \midrule
    \multicolumn{2}{c}{Audio-visual SRE} & $03.70$ & $0.141$ & $0.166$ ~~~             \\
    \multicolumn{2}{c}{\bf with VFNet} & $\textbf{03.20}$ & $\textbf{0.141}$ & $\textbf{0.141}$ ~~~             \\
    \toprule
    \hline
    \multicolumn{5}{c}{\textbf{Evaluation Set}} \\
    \midrule
    %\multicolumn{2}{c}{\textbf{System}} & \textbf{MinDCF} & \textbf{ActDCF} & \textbf{EER (\%)} \\ 
  %  \midrule
    \multicolumn{2}{c}{Speaker Recognition} & $06.36$ & $0.326$ & $0.339$ ~~~             \\
    \multicolumn{2}{c}{with VFNet} & $05.79$ & $0.317$ & $0.320$ ~~~             \\
    \midrule
    \multicolumn{2}{c}{Face Recognition} & $01.77$ & $0.074$ & $0.098$ ~~~             \\
    \multicolumn{2}{c}{with VFNet} & $01.66$ & $0.073$ & $0.094$ ~~~             \\
    \midrule
    \multicolumn{2}{c}{Audio-visual SRE} & $01.33$ & $0.050$ & $0.068$ ~~~             \\
    \multicolumn{2}{c}{\bf with VFNet} & $\textbf{01.11}$ & $\textbf{0.049}$ & $\textbf{0.062}$ ~~~             \\
    \bottomrule
  \end{tabular}}
  }
   % \vspace{-4mm}
\end{table}

\section{Conclusions}
\label{conc}

In this work, we propose a novel framework for audio-visual speaker recognition with cross-modal discrimination network. The VFNet based cross-modal discrimination network finds the relation between a given pair of human voice and face to generate a confidence score if they correspond to the same person. While VFNet can perform comparable to the existing state-of-the-art cross-modal verification systems, the proposed framework of audio-visual speaker recognition with VFNet  outperforms the baseline audio-visual system. This highlights the importance of cross-modal verification, in other words, the relation between audio and visual cues for audio-visual speaker recognition.

%In this paper, we propose a novel audio-visual speaker recognition framework: VFNet aided audio-visual SRE, which utilizes the cross-modal verification system to support the original audio-visual SRE framework by adding the discovered information between voice and face. The results prove that our cross-modal verification model, VFNet, can effectively improve the final audio-visual speaker recognition results. Meanwhile, VFNet can achieve the comparable existing state-of-the-art cross-modal verification and matching performance. This paper reveals the application value of cross-modal system in other mainstream recognition tasks like speaker recognition, face recognition and audio-visual speaker recognition.

%\newpage
\section{Acknowledgements}
This research work is partially supported by Programmatic Grant No. A1687b0033 from the Singapore Government's Research, Innovation and Enterprise 2020 plan (Advanced Manufacturing and Engineering domain), and in part by Human-Robot Interaction Phase 1 (Grant No. 192 25 00054) from the National Research Foundation, Prime Minister's Office, Singapore under the National Robotics Programme.

%, and is also part of a research collaboration with Kriston AI Lab.

%\newpage
\balance
\bibliographystyle{IEEEtran}
\bibliography{mybib}
\end{document}